\documentclass[aip,apl,preprint,amsmath,amssymb]{revtex4-2}

\usepackage{graphicx}

\begin{document}

\title{Nanowrinkle Waveguide in Graphene for Enabling Secure Dirac Fermion Transport}

\author{Seunghyun Jun}
\affiliation{Department of Physics Education, Chosun University, Gwangju 61452, Republic of Korea}

\author{Myung-Chul Jung}
\affiliation{Department of Physics Education, Chosun University, Gwangju 61452, Republic of Korea}
\affiliation{Institute of Well-Aging Medicare \& Chosun University G-LAMP Project Group, Chosun University, Gwangju, 61452, Republic of Korea}

\author{Nojoon Myoung}
\email{nmyoung@chosun.ac.kr}
\affiliation{Department of Physics Education, Chosun University, Gwangju 61452, Republic of Korea}
\affiliation{Institute of Well-Aging Medicare \& Chosun University G-LAMP Project Group, Chosun University, Gwangju, 61452, Republic of Korea}

\date{\today}

\begin{abstract}
Localized states in graphene have garnered significant attention in quantum information science due to their potential applications. Despite graphene's superior transport and electronic properties compared to other semiconductors, achieving nanoscale confinement remains challenging due to its gapless nature. In this study, we explore the unique transport properties along nanowrinkles in monolayer graphene. We demonstrate the creation of a one-dimensional conduction channel by alternating pseudo-magnetic fields along the nanowrinkle, enabling ballistic Dirac fermion transport without leakage. This suggests a feasible method for secure quantum information transfer over long distances. Furthermore, we extend our analysis to bent nanowrinkles, showcasing well-guided Dirac fermion propagation unless the bent angle is sufficiently large. Our demonstration of the nanowrinkle waveguide in graphene introduces a novel approach to controlling Dirac fermion transport through strain engineering, for quantum information technology applications.
\end{abstract}

\maketitle

\section{Introduction}
Graphene, a two-dimensional material composed of a single layer of carbon atoms arranged in a hexagonal lattice, has attracted significant attention since its discovery due to its remarkable electronic, mechanical, and thermal properties~\cite{novoselov2005,Geim2007}. Graphene exhibits high charge carrier mobility, exceptional mechanical strength, and superior thermal conductivity~\cite{novoselov2005,Geim2007,Changgu2008,Neto2009,Ando2009,Zhai2011}. These properties make graphene an ideal candidate for various applications, from flexible electronics to high-frequency transistors~\cite{}. In addition, the distinctive electronic structure of graphene, characterized by its Dirac fermions, makes it a promising material for quantum information science and valleytronics~\cite{ novoselov2005,Geim2007,Changgu2008,Gusynin2005, Neto2009, Ando2009, Zhang2005,Zhai2011}.

However, contrary to expectations, the energy gap-free nature of graphene meant that the current could not be controlled in the same way as in many conventional semiconductor devices. Various methods were proposed to open an energy gap in graphene, but the electrical properties of graphene get worse than those of pristine graphene~\cite{ Neto2009, Ando2009}. To overcome this, mechanical deformation of graphene was utilized. In actual synthesis, graphene is mechanically deformed in a variety of ways. New quantum states can be tuned by structural modifications, it could lead to a variety of quantum devices using graphene with superior transport properties~\cite{ Zhao2009, Liu2007, Fujita2010, Zhai2011}. Recently, many efforts have been made to create quantum dots on graphene and utilize them for applications such as quantum information devices or quantum information transportation~\cite{ Trauzettel2007, yamamoto2012electrical, Tian2018, Eich2018, Banszerus2022, Banszerus2023, Garreis2024, Henna2020}. In this paper, we focus on the quantum information transportation in strained graphene.

Lattice deformation of graphene to enhance or manipulate its electrical properties has been widely studied~\cite{Wei2018,Son2010,Si2016,Pereira2009, Polyzos2015,Kim2012,Cao2020,Ghorbanfekr2017}. As a result of this lattice deformation, it has been found that the Dirac fermions behave as if they feel a magnetic field, which is called pseudomagnetic field (PMF)~\cite{ Levy2010, hsu2020nanoscale, Low2010, Jia2019, banerjee2020strain, Zhang2014}. The PMF in nanobubble or nanowinkle can affect the momentum of electrons or holes in graphene, leading to interesting physical phenomena such as the formation of the zero-field Landau level or the valley Hall effect~\cite{ hsu2020nanoscale, Nigge2019, li2020valley, liu2022realizing, Ren2022}. 

These strain-induced fields create new pathways for electronic transport, especially in nanowinkle, allowing the formation of one-dimensional (1D) conducting channels~\cite{ lim2015structurally, liu2022realizing, Zhang2014}. These channels are particularly interesting for quantum information applications because they can guide Dirac fermions with minimal scattering and decoherence~\cite{}. The ability to control the propagation of Dirac fermions through strain engineering opens new avenues for the development of graphene-based quantum devices, including waveguides and logic gates, which are essential for building scalable quantum networks.

In this paper, we demonstrate a strain-engineered graphene waveguide for use as a quantum information device, and identify the waveguide mode by the PMF due to a nanowrinkle using the tight-binding model. Our results show that the waveguide enables quantum information transfer through 1D conduction channels. We further show that Dirac fermions are well guided along a bent nanowrinkle, examining the effects of the bent angle on its transport properties. 

The manuscript is organized as follows. In Sec. \ref{sec:model}, we introduce the theoretical frameworks of the system, addressing the characteristics of the strain-induced PMF and methodologies for both analytic analysis and numerical calculations. In Sec. \ref{sec:waveguide}, we demonstrate the occurrence of the waveguide mode by presenting the band structures and transport properties through the nanowrinkle. We show that Dirac fermions are indeed well-guided along the nanowrinkle waveguide, exhibiting perfect transmission. Furthermore, in Sec. \ref{sec:bent}, we explore the effects of waveguide modulation on Dirac fermion transport by considering bent nanowrinkles. We reveal that bent nanowrinkles can also host waveguide modes, indicating the potential application of the nanowrinkle waveguide as a building block for quantum devices or circuits. Finally, in Sec. \ref{sec:conclusion}, we summarize our findings and discuss the implications of our results for future research and potential applications in quantum information technology.

\section{Results}

\subsection*{Strain-induced pseudo-magnetic fields}\label{sec:model}

\begin{figure}[tpbt!]
    \centering
    \includegraphics[width=\linewidth]{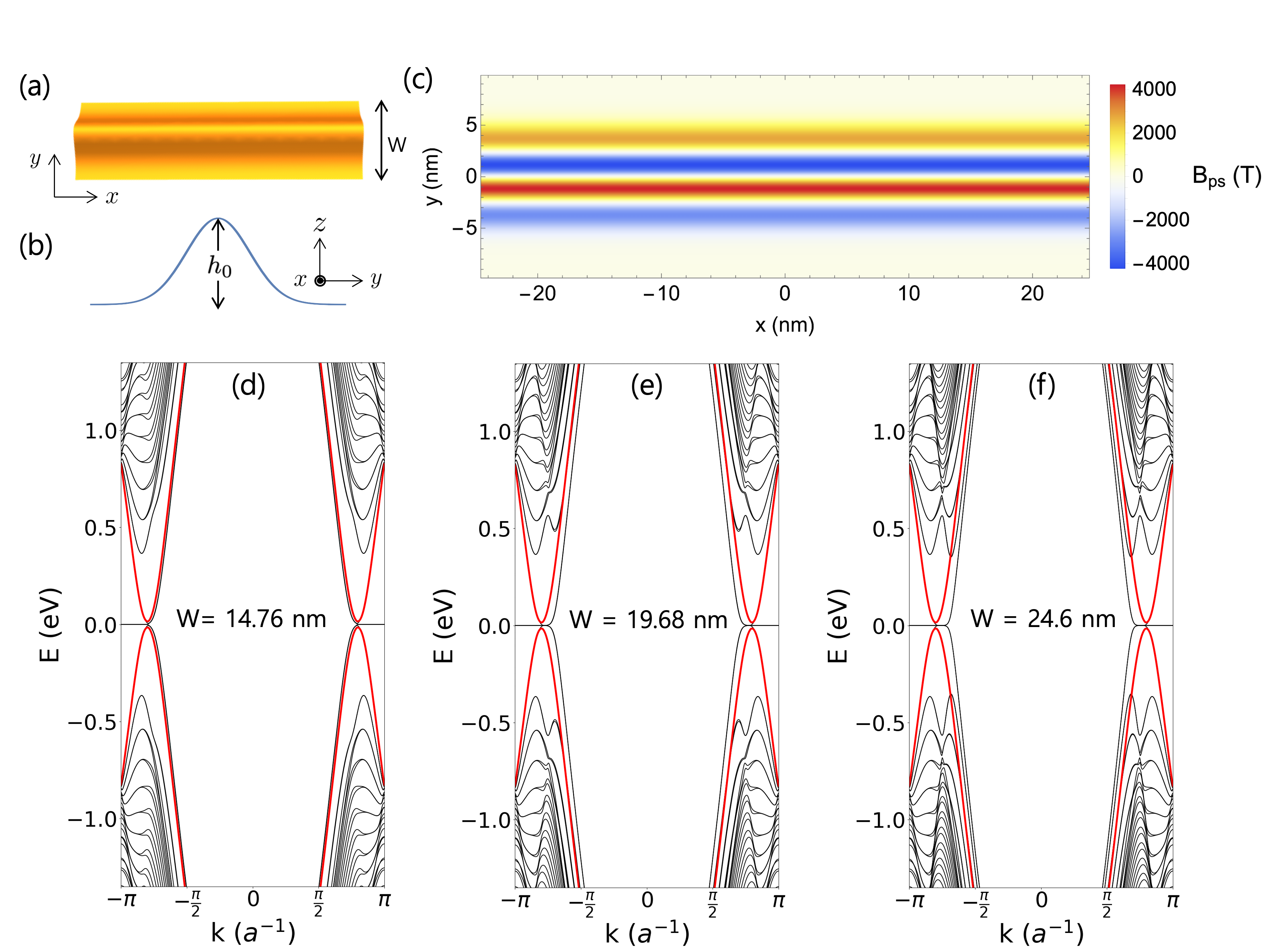}
    \caption{(a) Schematic of a nanowrinkle along a graphene nanoribbon with width $W$. (b) Side view of the nanowrinkle modeled as a Gaussian function with maximum height $h_{0}$. (c) Pseudo-magnetic field (PMF) profile calculated from the low-energy effective Hamiltonian. (d-f) Energy bands of the graphene nanoribbon subjected to the PMF due to the nanowrinkle, shown for different ribbon widths. The strain-induced waveguide modes in the band structures are highlighted in red, indicating their invariance with respect to ribbonwidth variation. The nanowrinkle parameters used are $h_{0} = 3.69~\mathrm{nm}$ and $\sigma = 2.46~\mathrm{nm}$.}
    \label{fig:model}
\end{figure}

Figure \ref{fig:model}(a) and (b) depict a theoretical model of the nanowrinkle, defined by a Gaussian function\cite{zhai2019electron}:
\begin{align}
    z\left(\vec{r}\right)=h_{0}e^{-\frac{\left(y-y_{0}\right)^{2}}{2\sigma^{2}}},
\end{align}
where $h_{0}$ represents the maximum vertical deformation of the nanowrinkle and $\sqrt{2}\sigma$ denotes the standard deviation of the Gaussian profile, indicating the lateral size of the nanowrinkle. These nanowrinkles can form accidentally during the transfer process onto a substrate, or can be intentionally created via various strain engineering techniques\cite{kang2020role,lim2015structurally,pacakova2017mastering}. Due to the local strain accompanied by the lattice deformation of the nanowrinkle, strain-induced gauge fields emerge in the strained region:
\begin{align}
    \vec{A}_{ps}\left(\vec{r}\right)=-\frac{\sqrt{3}\beta\hbar}{ea_{0}}\frac{h_{0}^{2}}{2\sigma^{4}}y^{2}e^{-\left(y-y_{0}\right)^{2}/\sigma^{2}}\hat{x},
\end{align}
assuming the zigzag-edge orientation is along the $x$-axis. Here, $\beta=3.37$ and $a_{0}=0.246~\mathrm{nm}$. Consequently, the PMF due to the nanowrinkle is obtained by $\vec{B}_{ps}=\vec{\nabla}\times\vec{A}_{ps}$:
\begin{align}
    \vec{B}_{ps}=\frac{\beta\hbar}{\sqrt{3}ea_{0}}\frac{h_{0}^{2}}{\sigma^{6}}y\left(\sigma^{2}-y^{2}\right)e^{-\left(y-y_{0}\right)^{2}/\sigma}\hat{z}.\label{eq:PMF}
\end{align}

As depicted in Fig. \ref{fig:model}(c), the nanowrinkle can host three `snake-orbit' conducting channels\cite{hsu2020nanoscale,banerjee2020strain} along the `zero'-PMF lines at $y=0$ and $\pm\sigma$. To confirm the existence of the strain-induced conducting channels in the nanowrinkle, we examine the band structures of zigzag graphene nanoribbons subjected to nanowrinkles, varing the ribbonwidth. Figures \ref{fig:model}(d-f) clearly show a robust band, highlighted in red, which remains unchanged with respect to the system size. This band is purely associated with the presence of nanowrinkle. Notably, the degeneracy of the strain-induced waveguide mode is two-fold due to the preserved pseudospin degeneracy of graphene.

To examine the transport properties of the nanowrinke waveguide in graphene, we investigate transport phenomena through graphene subjected to a nanowrinkle-induced PMF using numerical calculation with the S-matrix approach based on the tight-binding Hamiltonian of graphene:
\begin{align}
    H=\sum_{\left<i,j\right>}\left(t_{ij}a^{\dagger}_{i}b_{j}+\mathrm{h.c.}\right),
\end{align}
with the modified hopping term due to lattice deformation:
\begin{align}
    t_{ij}=t_{0}e^{-\beta\left(\frac{d_{ij}}{a_{0}}-1\right)},
\end{align}
where $t_{0}=2.7~\mathrm{eV}$, and $d_{ij}$ is the distance between adjacent atomic sites in the strained graphene lattice. The quantum conductance through the system is obtained using the Landauer-B\"{u}ttiker formalism:
\begin{align}
    G\left(E\right)=\frac{2e^{2}}{h}\sum_{\alpha,\beta}T_{\alpha\beta}\left(E\right),\label{eq:Landauer}
\end{align}
where $T_{\alpha,\beta}$ is the transmission function from a lead$\beta$ to a lead$\alpha$ at a given energy $E$, numerically acquired from the S-matrix of the systems. We take into account the spin degeneracy as a factor 2 in Eq. (\ref{eq:Landauer}).

\subsection*{Graphene nanowrinkle waveguide}\label{sec:waveguide}

\begin{figure}[htpb!]
    \centering
    \includegraphics[width=\linewidth]{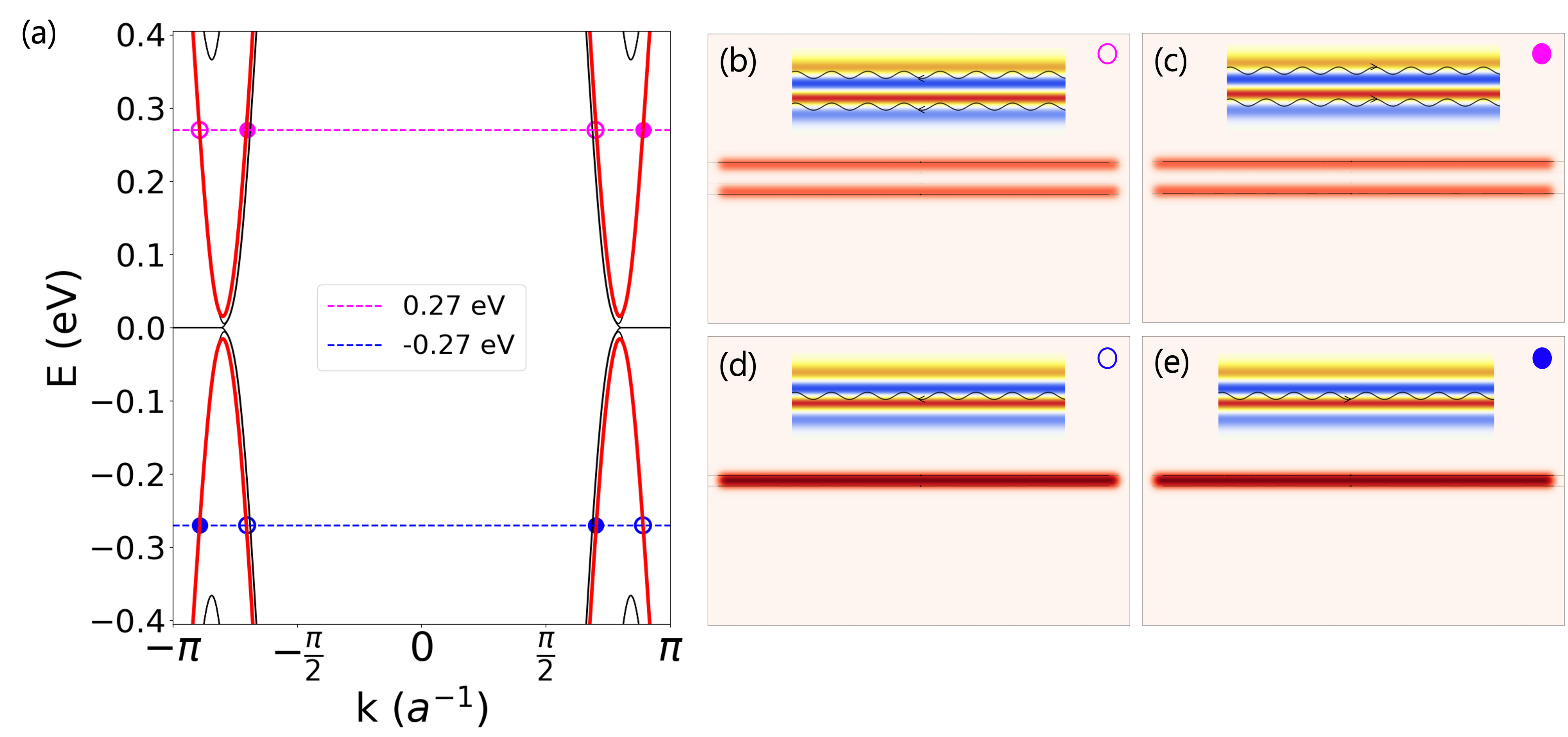}
    \caption{(a) Zoom-in band structure from Fig. \ref{fig:model}(f). (b-d) Probability current density maps through the system. Each current map corresponds to the left/right-propagating mode in the case of electron/hole transport, respectively denoted as open/solid color circles in (a). Insets in (b-d) are the schematic diagrams of the allowed snake-orbit channels, respectively corresponding to the current maps. Note that all maps are normalized to the same values.}
    \label{fig:channel}
\end{figure}

As aforementioned, the snake-orbit channels can be formed at $y=0$ and $\pm\sigma$, with the propagation directions along the channel at $y=0$ being opposite to those at $y=\pm\sigma$. This spatial distinction of the right- and left-propagating waveguide modes arises from the opposite signs of the PMF profiles near the interfaces $y=0$ and $\pm\sigma$. Figure \ref{fig:model}(c) displays the PMF profile obtained from the low-energy effective Hamiltonian model in continuum limit:
\begin{align}
    H&=v_{F}\hbar\vec{\sigma}\cdot\left(\vec{k}+se\vec{A}_{ps}\right)\nonumber\\
    &=v_{F}\hbar\left[\sigma_{x}\left(k_{x}+seA_{ps,x}\right)+\sigma_{y}k_{y}\right],\label{eq:effH}
\end{align}
obeying the Dirac equation, $H\psi=E\psi$, with its eigenenergy $E=\hbar v_{F}\sqrt{\left(k_{x}+seA_{ps,x}\right)^{2}+k_{y}^{2}}$, where $v_{F}\simeq 10^{6}~\mathrm{ms}^{-1}$ is the Fermi velocity of graphene. As shown in Fig. \ref{fig:model}(d-f), the energy of the system is preserved with momentum inversion in the $x$ direction, i.e., $E\left(-k_{x}\right)=E\left(k_{x}\right)$. Thus, the strain-induced PMF $\vec{B}_{ps}$ should be accordingly reversed as the propagation direction changes, i.e., $s=\mathrm{sign}\left(k_{x}\right)$ from Eq. (\ref{eq:effH}). Indeed, as displayed in Fig. \ref{fig:channel}(b) and (c), the right-propagating electron moves along the channels at $y=\pm\sigma$, and the left-propagating electron also moves along the same channels, since it feels the reversed PMF compared to that for the left-propagating electron. On the other hands, for hole transport, the conducting channels in the nanowrinkle waveguide are formed at $y=0$, as shown in Fig. \ref{fig:channel}(d) and (e).

\begin{figure*}[htpb!]
    \centering
    \includegraphics[width=\linewidth]{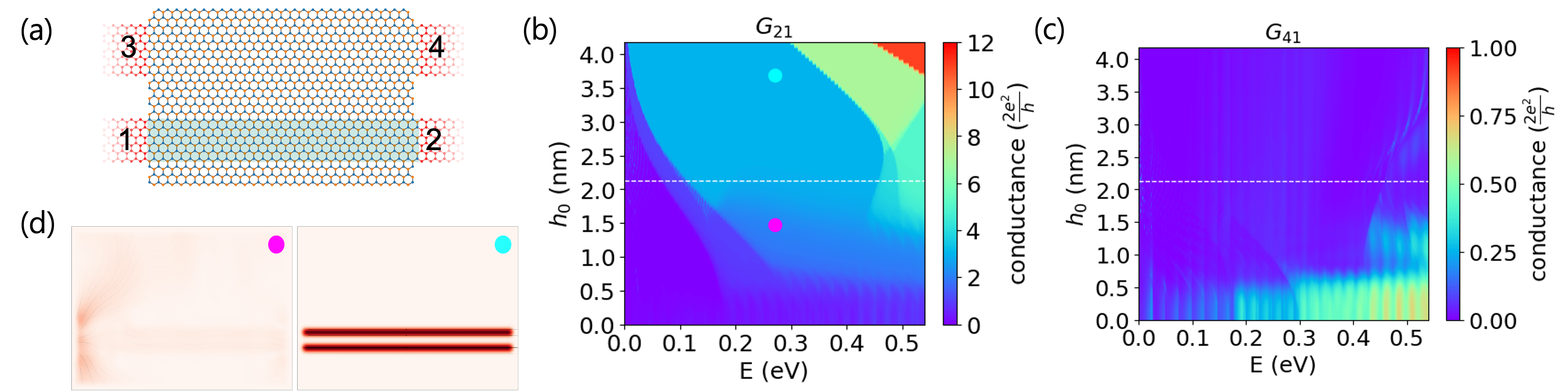}
    \caption{(a) Schematics of the system with four leads. The shaded region indicates the existence of a nanowrinkle connecting two leads, lead1 and lead2. (b) and (c) Conductance spectra $G_{21}$ and $G_{41}$ versus Dirac fermion energy $E$ and the nanowrinkle height $h_{0}$. The dashed lines represent the critical nanowrinkle height to secure the waveguide mode, $2.13~\mathrm{nm}$. (d) Probability current density maps at a given energy $E=0.27~\mathrm{eV}$, for different $h_{0}$ values, $1.476~\mathrm{nm}$ and $3.69~\mathrm{nm}$, respectively denoted by colored circles in (b). A parameter used for the calculations is $\sigma=2.46~\mathrm{nm}$.}
    \label{fig:waveguide}
\end{figure*}

We now examine the transport properties of the graphene nanowrinkle waveguide to consider its potential application for quantum information transfer. First, we need to be confirm whether Dirac fermions transport is truly guided by the PMF due to the nanowrinkle. To investigate this, we introduce a four-terminal graphene device, where a straight nanowrinkle is placed between two leads, labeled as lead1 and lead2, as depicted in Fig. \ref{fig:waveguide}(a). If the nanowrinkle indeed plays a role of guiding Dirac fermions, incident Dirac fermions from the lead1 will not be transmitted to the lead3 and the lead4, but will be entirely collected by the lead2.

The calculated results of the conductance, $G_{21}$ and $G_{41}$, are shown in Fig. \ref{fig:waveguide}(b) and (c). For small $h_{0}$, i.e., for a relatively week PMF, the conductance spectra $G_{21}$ exhibit the characteristic step-wise increments as a function of Dirac fermion energy, reflecting the quasi-1D nature of the zigzag graphene nanoribbon. In this case, it is clear that $G_{41}$ spectra do not vanish as shown in Fig. \ref{fig:waveguide}(c), because Dirac fermions are not guided but spread out over the system. On the other hand, as $h_{0}$ increases, $G_{41}$ spectra become substantially suppressed, and the step-wise behavior of $G_{21}$ spectra changes compared to the $h_{0}=0$ case. Interestingly, the new conductance plateaus for large $h_{0}$ values are in good agreement with the band structure displayed in Fig. \ref{fig:model}(e).

Although the contribution of the waveguide mode to the conductance plateau in $G_{21}$ cannot be perfectly resolved, we can directly confirm the formation of the nanowrinkle waveguide by observing Dirac fermion transport phenomena for a large $h_{0}$ value. Figure \ref{fig:waveguide}(d) shows a comparison of the current densities between two cases; non-guided and guided transport of Dirac fermions, which are respectively denoted by different color circles in Fig. \ref{fig:waveguide}(b). Indeed, for the small $h_{0}$ case, Dirac fermions are spread out over the entire system, whereas for the large $h_{0}$ case, Dirac fermions are perfectly guided inside the strained region where the nanowrinkle is created.

The waveguide mode from the band structures depicted in Fig. \ref{fig:model}(e) is doubly degenerate, leading to a conductance contribution as much as $2\times2e^{2}/h$. However, the magnitude of the conductance plateau in Fig. \ref{fig:waveguide}(b) is found to be $3\times2e^{2}/h$, because the zero-energy mode of the zigzag graphene nanoribbon contributes an additional $2e^{2}/h$ to the waveguide mode.

Moreover, it is worth considering how large $h_{0}$ needs to be to create a waveguide. To address this question, we now conduct a theoretical analysis using a semiclassical approach based on the Dirac equation. Though the PMF profile due to the nanowrinkle is non-uniform in the $y$ direction, we can simplify the system by considering a staggered PMF profile, where each region is under a uniform PMF. In this study, the reduced PMF strengths for a given nanowrinkle with $h_{0}=3.69~\mathrm{nm}$ and $\sigma=2.46~\mathrm{nm}$ are calculated to be $2600~\mathrm{T}$ and $1310~\mathrm{T}$, respectively. The corresponding magnetic lengths are found to be $0.5~\mathrm{nm}$ and $0.7~\mathrm{nm}$, respectively. Since the magnetic length indicates the radius of the snake-orbit trajectory, these length scales correspond to the spectral width of the interface channels quantum mechanically. The distance between channels in the nanowrinkle waveguide is given by $\sigma=2.46~\mathrm{nm}$, which is much larger than the magnetic lengths calculated for our nanowrinkle waveguide.

Similarly, we can evaluate the corresponding magnetic lengths for $h_{0}=0.5~\mathrm{nm}$, $3.75~\mathrm{nm}$ and $5.24~\mathrm{nm}$, which are larger than $\sigma/2$. Thus, for a small $h_{0}$ value, the resulting PMF strength is insufficient to distinguish two waveguide channels, resulting in the significant admixture of two snake-orbit states. This analysis provides a critical value of $h_{0}$ for a nanowrinkle to be considered a good waveguide, determined by finding the required $h_{0}$ value for the magnetic length is smaller than $\sigma/2$. After some algebra, we obtain the critical $h_{0}$ value of $2.13~\mathrm{nm}$ for the given $\sigma=2.46~\mathrm{nm}$. As shown in Fig. \ref{fig:waveguide}(b) and (c), the $G_{41}$ spectra indeed vanish for $h_{0}>2.13~\mathrm{nm}$, as the conducting channels start being spatially resolved by the sufficiently strong PMF. 

\subsection*{Bent waveguide for strain engineering}\label{sec:bent}

\begin{figure*}[htpb!]
    \centering
    \includegraphics[width=\linewidth]{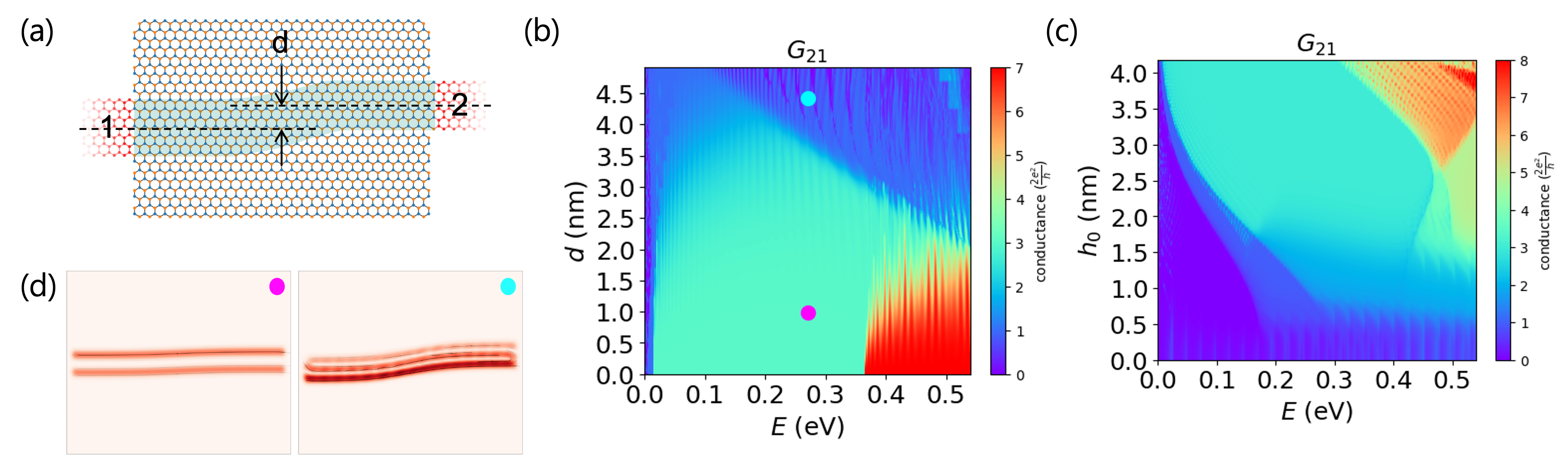}
    \caption{(a) Schematic view of the system with a bent nanowrinkle. The shaded region indicates the existence of the bent nanowrinkle connecting two leads, lead1 and lead2. (b) Conductance spectra versus Dirac fermion energy $E$ and the separation between two leads $d$, for a given $h_{0}=3.69~\mathrm{nm}$. (c) Conductance spectra versus Dirac fermion energy $E$ and the nanowrinkle height $h_{0}$, for a given $d=0.984~\mathrm{nm}$. (d) Probability current density maps at a given energy $E=0.27~\mathrm{eV}$, for different $d$ values, $0.984~\mathrm{nm}$ and $4.428~\mathrm{nm}$, respectively denoted by colored circles in (b). A parameter used for the calculations is $\sigma=2.46~\mathrm{nm}$.}
    \label{fig:bent}
\end{figure*}

So far, we have examine the properties of the nanowrinkle waveguide, which is straight. In terms of strain engineering, the effects of bending on the nanowrinkle waveguide can be important, especially when designing sophisticated waveguide devices. For instance, electronic waveguides made of semiconductor have been used to fabricate quantum devices like quantum interferometry or flying qubit devices\cite{yamamoto2012electrical,bautze2014theoretical,takada2019sound,roussely2018unveiling}. The bent nanowrinkle waveguide in our study is displayed in Fig. \ref{fig:bent}(a), which its conformal shape formulated by
\begin{align}
    z\left(\vec{r}\right)=h_{0}e^{-\left[y-f\left(x\right)\right]^{2}/2\sigma^{2}},
\end{align}
with a given functional landscape:
\begin{align}
    f\left(x\right)=d\tanh{\left(\frac{x}{\xi}\right)},
\end{align}
where $d$ is the separation between the centers of the two leads, and $\xi$ indicates the steepness of the bending. The bent angle is defined by using $d$ and $\xi$:
\begin{align}
    \theta=\tan^{-1}{\left(\frac{d}{\xi}\right)}.
\end{align}
Here, we consider an appropriate value of $\xi=24.6~\mathrm{nm}$, which is within a practical device scale.

First, we examine the transport properties of the bent waveguide to determine whether Dirac fermion transport is also well guided along the bent waveguide. The results are plotted in Fig. \ref{fig:bent}(b). For small $d$, the conductance spectra exhibit a clear plateau at $G=3\times2e^{2}/h$, indicating the bent nanowrinke also hosts the waveguide mode, as expected. As $d$ increases, however, the conductance plateau is reduced to $G=2e^{2}/h$. This reduction can be understood by examining how the current densities flow throughout the system, as shown in Fig. \ref{fig:bent}(d). For the small bent angle $2.3^{\circ}$ (denoted by the magenta circle), Dirac fermions propagate well along the bent waveguide, similary to the straight nanowrinkle case. On the other hand, for the sufficiently large bent angle $10.2^{\circ}$, the net flux of the probability current via the waveguide modes is cancelled out due to the coexistence of thr right- and left-propagating current flows. (Note that the right-propagating mode delivers the same conductance contribution as the left-propagating mode, even if the right-propagating mode is spatially separated at $y=\pm\sigma$, as aforementioned.) Despite of the washed-out waveguide modes, there still remains the $2e^{2}/h$ conductance contribution from the zero-energy mode of the zigzag graphene nanoribbon.

To confirm the existence of the bent waveguide, we also examine the conductance spectra through the given bent nanowrinkle with the given bent angle $2.3^{\circ}$. Figure \ref{fig:bent}(c) presents the resulting conductance map through the bent nanowrinkle as functions of $E$ and $h_{0}$. It is evident that the transport properties of the bent nanowrikle are almost identical to the straight nanowrinkle, clearly exhibiting the $3\times 2e^{2}/h$ conductance plateau.

In addition, one can observe that the conductance spectra shown in Fig. \ref{fig:bent}(b) and (c) exhibit a fluctuating nature as a function of $E$. It is also worth noting that the reduction in the conductance through the bent waveguide is due to the emergent backscattering. The presence of the left-propagating mode in Fig. \ref{fig:bent}(d) is allowed if the right-propagating electron from the left lead (lead1) is backscattered. Such back-and-forth propagation along the nanowrinkle waveguide can lead to Fabry-Perot interference in the conductance spectra, resulting in the fluctuation in the conductance.

\section{Discussion}\label{sec:conclusion}

In this work, we have demonstrated that a nanowrinkle in graphene can be used for a Dirac fermion waveguide due to the strain-induced pseudo-magnetic field (PMF). The waveguide mode in the nanowrinkle is formed at the zero-PMF line, contributing a conductance of $4e^{2}/h$. Interestingly, we have further revealed that the nanowrinkle waveguide can be engineered by modifying its shape. Dirac fermion transport remains well-guided along the bent waveguide, if the bent angle is less than a certain value.

Our graphene nanowrinkle waveguide offers several advantageous for applications in quantum information technologies. First, we can realize the control of Dirac fermion transport in monolayer graphene without opening a gap, thereby preserving its transport properties. With the rapid progress in strain engineering techniques for graphene and 2D materials, such strain-induced nanostructures in graphene are now feasible for experimental studies\cite{zhai2019electron,hsu2020nanoscale,colangelo2018controlling,pacakova2017mastering,yan2013strain,oliveira2015crystal,li2020valley,liu2022realizing}. Second, our nanowrinkle waveguide inherently minimize information loss due to leackage, as the snake-orbit interface channel is basically topologically protected\cite{liu2015snake,myoung2017conductance}. Unless the PMF profile of the nanowrinkle is significanlty altered, the existence of the conducting channels between regions of opposite-sign PMF is robust against common disorders such as charge inhomogeneity.

In conclusion, our results suggest that nanowrinkles in graphene can serve as a building block for graphene-based quantum information technology, such as flying qubits.

\section{Methods}

Finite-size tight-binding Hamiltonian construction and numerical calculations of the transport properties of the graphene nanowrinkle waveguide are performed using the \textsc{kwant} package with the \textsc{mumps} routines\cite{groth2014kwant,amestoy2001fully}.

\section*{Data availability}

The data supporting the findings of this study are available from the corresponding authors upon reasonable request.

\section*{Code availability}

The codes used in this study are available from the corresponding author on reasonable request.

\section*{References}
\bibliography{NWGra}

\begin{acknowledgments}
This work was supported from the National Research Foundation of Korea(NRF) grant funded by the Ministry of Education(No. RS-2023-00285353) and Ministry of Science and ICT(No. NRF-2022R1F1A1065365). The authors are grateful to Prof. Hee Chul Park for his insightful discussions and contributions to this work.
\end{acknowledgments}

\section*{Author information}
\subsection*{Contributions}
SJ performed the numerical calculations. NM designed and oversaw the project. MCJ and NM contributed to writing the manuscript.
\subsection*{Corresponding author}
Correspond to Nojoon Myoung.

\section*{Ethics declarations}
\subsection*{Competing interests}
The authors declare no competing interests.

\end{document}